\documentclass[12pt]{spieman}  
\usepackage{amsmath,amsfonts,amssymb}
\usepackage{graphicx}
\usepackage{setspace}
\usepackage{tocloft}

\usepackage{textcomp} 

\usepackage{float}
\usepackage[utf8]{inputenc}
\usepackage[T1]{fontenc}
\usepackage{mathptmx}

\title{Resonant absorption in multilayer quantum-well and quantum-dot solar cells}

\author[a,*]{Maxime Giteau}
\author[a]{Yusuke Oteki}
\author[b]{Kento Kitahara}
\author[a]{Naoya Miyashita}
\author[a]{Ryo Tamaki}
\author[a,b]{Yoshitaka Okada}
\affil[a]{Research Center for Advanced Science and Technology, The University of Tokyo, Meguro, Tokyo 153-8904, Japan}
\affil[b]{Department of Electrical Engineering and Information Systems, The University of Tokyo, Bunkyo, Tokyo 113-8656, Japan}

\cftpagenumbersoff{figure}
\cftpagenumbersoff{table} 
\begin{document}

\maketitle

\begin{abstract}

Epitaxially-grown quantum well and quantum dot solar cells suffer from weak light absorption, strongly limiting their performance.
Light trapping based on optical resonances is particularly relevant for such devices to increase light absorption and thereby current generation. 
Compared to homogeneous media, the position of the quantum layers within the device is an additional parameter that can strongly influence resonant absorption.
However, this effect has so far received little attention from the photovoltaic community.
In this work, we develop a theoretical framework to evaluate and optimize resonant light absorption in a thin slab with multiple quantum layers.
Using numerical simulations, we show that the position of the layers can make the difference between strong absorption enhancement and completely suppressed absorption, and that an optimal position leads to an absorption enhancement twice larger than average.
We confirm these results experimentally by measuring the absorption enhancement from photoluminescence spectra in InAs/GaAs quantum dot samples.
Overall, this work provides an additional degree of freedom to substantially improve absorption, encouraging the development of quantum wells and quantum dots-based devices such as intermediate-band solar cells.

\end{abstract}


{\noindent \footnotesize\textbf{*}Maxime Giteau,  \linkable{maxime.giteau.pro@gmail.com} }

\begin{spacing}{2}   

\section{Introduction}

Quantum layers (QLs), like epitaxially-grown quantum wells (QWs) and quantum dots (QDs), are broadly used in opto-electronic devices such as lasers, photodetectors and solar cells~\cite{ustinov_quantum_2003,schneider_quantum_2007,nozik_quantum_2002}.
For photovoltaics, they are investigated to extend the absorption edge of a given material~\cite{das_improved_2018}, as well as for high-effiency concepts like intermediate-band \cite{luque_increasing_1997,okada_recent_2013}, multi-exciton generation~\cite{nozik_quantum_2002,semonin_peak_2011} and hot-carrier solar cells~\cite{nguyen_quantitative_2018,harada_hot-carrier_2019}.

Strong light absorption is critical for the performance of solar cells.
However, a single QL absorbs only a small fraction of the incident light. 
To increase optical absorption, one must increase the number of QLs~\cite{liu_high_2001,sugaya_ultra-high_2012} and/or enhance the absorption per layer through light trapping~\cite{mellor_feasibility_2014,zheng_recent_2016,johnson_advances_2005,ouellette_optical_2016,shoji_photoabsorption_2020,smith_inverted_2016,smith_development_2017,behaghel_absorption_2015,watanabe_thin-film_2014,elsehrawy_improved_2018,elsehrawy_guided-mode_2018,cappelluti_light-trapping_2018,aho_thin-film_2020}.
Depending on the application and materials required, too many QLs can lead to strain build-up and increased non-radiative recombinations which can reduce the efficiency of the device~\cite{takata_fabrication_2010}.
Light trapping is very attractive as it enables a larger absorption for a given number of QLs.
Resonant light trapping, which can be implemented when the thickness of the device is comparable with the wavelength of light (the device acts as a resonant cavity), has been widely applied to homogeneous absorbers, for example to achieve high broadband absorption in ultra-thin solar cell absorbers~\cite{yu_fundamental_2010,massiot_metal_2014,chen_199-efficient_2019}.
It has recently received significant attention for devices with QLs, where different types of resonances have been considered, including Fabry-Perot (FP) modes~\cite{behaghel_absorption_2015} and guided modes~\cite{watanabe_thin-film_2014,elsehrawy_improved_2018,elsehrawy_guided-mode_2018,cappelluti_light-trapping_2018}.
Still, light absorption in QL-based solar cells remains stongly insufficient, especially for enabling intermediate-band solar cells~\cite{mellor_feasibility_2014,ramiro_intermediate_2021},
and additional strategies are required to further increase absorption.

Resonant modes present an interesting feature for inhomogeneous media: the intensity of the field, and therefore the absorption enhancement, is not homogeneous in the cavity but oscillates between peaks and nodes depending on the position~\cite{saleh_fundamentals_2007}.
Therefore, as suggested by Behaghel \textit{et al.}, the absorption of QLs could be further increased by matching the position of the layers with the intensity peaks of resonant modes~\cite{behaghel_absorption_2015,behaghel_fabrication_2017}.
However, while the impact of the position of multiple QLs on the electric performances of solar cells has been considered extensively ~\cite{nelson_effect_1999,zhou_positioning_2010,kechiantz_impact_2015}, its impact on absorption has been largely overlooked, the absorber being usually considered as a homogeneous medium with an effective absorption coefficient~\cite{elsehrawy_guided-mode_2018}.

In this work, we introduce a theoretical framework to evaluate the absorption enhancement in devices with multiple QLs, depending on their position in a resonant cavity.
We derive analytical expressions for resonant absorption enhancement, and show that it can be twice larger in multilayer devices than in homogeneous media.
We then present numerical simulations of systems with single and multiple resonant modes, showing the crucial impact of the position of the QLs on their absorption. We also verify that an optimal QL position arrangement leads to an absorption twice stronger than a random position, in average.
Finally, we fabricate GaAs devices with multiple InAs QD layers and confirm, using photoluminescence, that resonances with optimized QD position show a larger absorption enhancement.

\section{Theory}

\subsection{Absorption enhancement in multiple quantum layers}

\begin{figure}
    \centering
    \includegraphics[width=.5\linewidth]{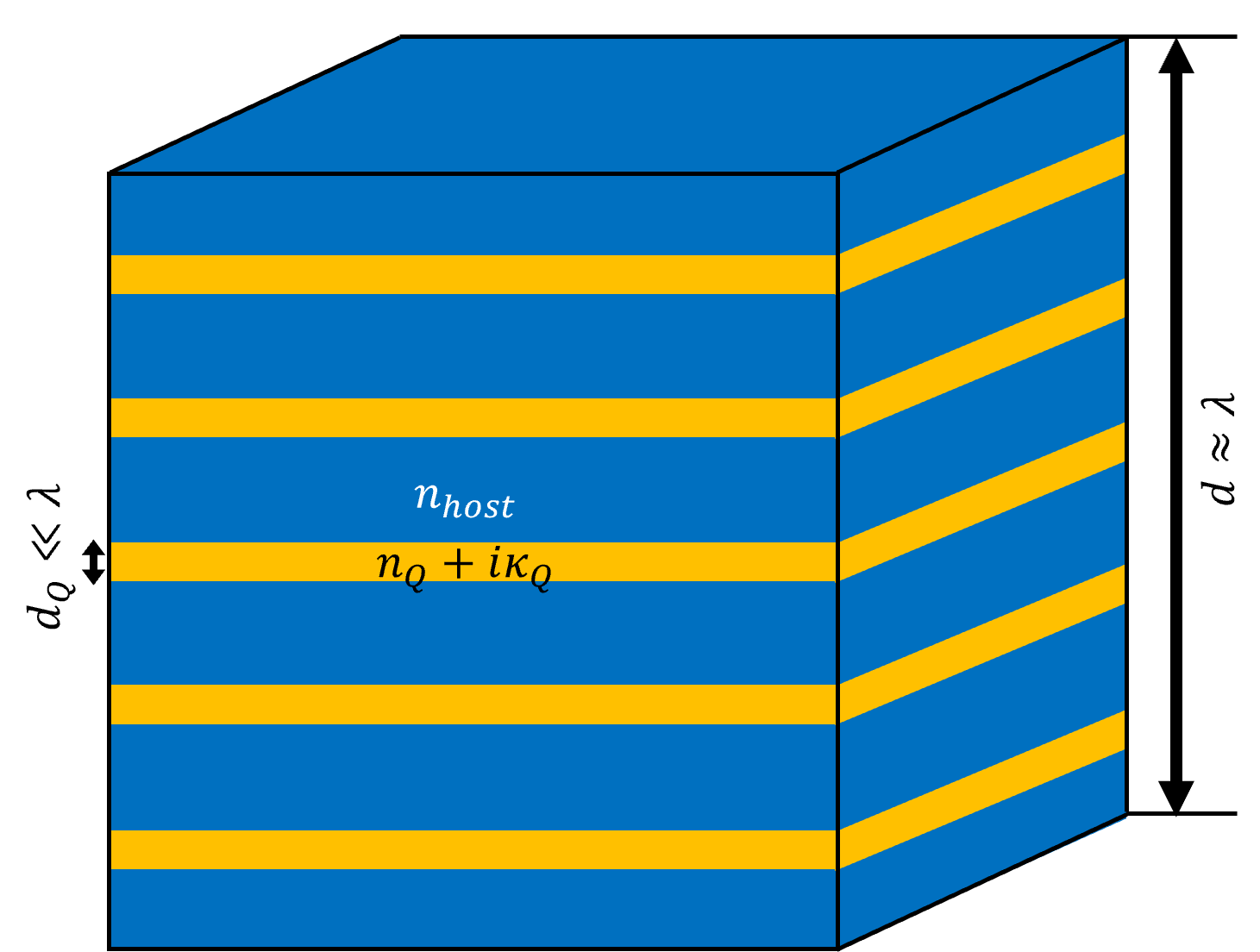}
    \caption{General multilayer structure considered in this work: an infinite slab of thickness $d\approx \lambda$ consisting of a host material real refractive index $n_{host}$ and $M$ QLs of complex refractive index $n_Q+ i\kappa_Q$ and thickness $d_Q\ll \lambda$ ($M=5$ in the figure). The slab is illuminated from the top at normal incidence by a monochromatic plane wave of wavelength $\lambda$.}
    \label{fig:Structure}
\end{figure}

We consider an infinite slab of thickness $d$ composed of a host material with refractive index $n_{host}$ and $M$ identical QLs of thickness $d_Q$ with complex refractive index $\tilde{n_Q}=n_Q+i\kappa_Q$ (Fig.\ref{fig:Structure}). 
We assume both materials are isotropic, homogeneous and non-magnetic, and the host material does not absorb light in the wavelength region of interest.
The slab is illuminated from the outside by a monochromatic plane wave of wavelength $\lambda$. 
For simplification, we consider normally incident light, although the discussion can be extended to oblique incidence.
The thickness $d$ of the cavity is comparable to the wavelength of light $\lambda$.
The thickness $d_Q$ of the QLs (typically a few nanometers) is much smaller than $\lambda$, so we can neglect reflection and scattering effects at the interfaces of the QLs.
The spectral power density absorbed in the QLs at a depth $z$ (in $\mathrm{W.cm^{-3}.nm^{-1}}$) can be directly related to the local intensity of the electric field $\mathbf{E}$~\cite{pettersson_modeling_1999}:

\begin{equation}
    p_{abs}(\lambda,z)=\frac{1}{2} c \ \varepsilon_0 \ n_Q(\lambda) \ \alpha_Q(\lambda) \ |\mathbf{E}(\lambda,z)|^2,
\end{equation}

\noindent where $c$ is the speed of light in vacuum, $\varepsilon_0$ is the dielectric permittivity of vacuum, and $\alpha_Q(\lambda)=4\pi \kappa_Q(\lambda)/\lambda$ is the absorption coefficient of the QLs.

Each QL $m$ is located between a depth $z_m$ and $z_m+d_Q$. 
The total absorption of the QLs is obtained by integrating the absorbed power over the thickness $d_Q$ of each QL, relative to the intensity of the incident field $\mathbf{E_{inc}}$ (we assume the incident medium is air):

\begin{equation}
    A(\lambda) = \alpha_Q(\lambda)\sum_{m}\int_{z_m}^{z_m+d_Q}{I(\lambda,z) dz},
    \label{eq:A_lambda}
\end{equation}

\noindent where $I(\lambda,z)$ is the normalized field intensity:

\begin{equation}
    I(\lambda,z) = n_Q(\lambda) \frac{|\mathbf{E}(\lambda,z)|^2}{|\mathbf{E_{inc}}(\lambda)|^2}.
\end{equation}

For simplification, we will consider very thin (few nm) QLs, such that the field can be considered homogeneous within a QL, leading to:

\begin{equation}
    A(\lambda) = \alpha_Q(\lambda) d_Q \sum_{m}{F_m(\lambda)},
    \label{eq:A_lambda_Fm}
\end{equation}

\noindent with

\begin{equation}
    F_m(\lambda) = I(\lambda,z_m).
    \label{eq:F_equal_I}
\end{equation}

$F_m$ corresponds to the absorption enhancement for the $m$-th QL relative to the single-pass absorption (assuming no front reflection and low absorption):

\begin{equation}
    A_{single}(\lambda) = \alpha_Q(\lambda) d_Q.
\end{equation}

Finally, we can express the average absorption enhancement $F$ (We call it "enhancement" since we focus on increasing absorption, although $F$ may be smaller than 1, as we will see later) in the QLs as:

\begin{equation}
    F(\lambda) = \frac{1}{M} \sum_{m}{F_m(\lambda)}.
    \label{eq:mean_enhancement}
\end{equation}

In the rest of this paper, we consider that host and QLs have the same real part of the refractive index $n = n_Q = n_{host}$, which is a fair approximation for III-V heterostructures.
If we assume a small absorption ($A < 0.1$) even in presence of light trapping, then we can consider that the intensity of the electric field is not modified by the absorption of the QLs, and the absorption enhancement $F_m$ can be estimated from the intensity distribution $I$ in absence of QLs, using Eq.~\ref{eq:F_equal_I}.
For stronger absorption, the field intensity will be attenuated in the presence of QLs.

\subsection{Maximum enhancement with quantum layers}
\label{s_theory}

Applying the temporal coupled-mode theory framework in a homogeneous cavity~\cite{haus_waves_1984,yu_fundamental_2010,collin_nanostructure_2014}, we show in Appendix~\ref{sec:AppendixA} that, in the limit of low absorption and assuming a perfect back mirror, the maximum light path enhancement at resonance wavelength $\lambda_0$ is:

\begin{equation}
    F(\lambda_0) \approx \frac{8}{\ln(1/R_f)}.
    \label{eq:Fhomo}
\end{equation}

\noindent where $R_f$ is the front surface reflection. 
We can generalize this result to non-homogeneous devices with multiple QLs in the following way.
The amplitude of a wave evolves as a sinewave as it propagates through space.
Therefore, the intensity $I$ of a resonant mode in a cavity oscillates with depth between 0 and its maximum value $I_{max}$, following a cosine square function whose average value is $I_{max}/2$.
In a homogeneous medium, absorption occurs as if the field had the same value $I_{max}/2$ everywhere.
As a result, the maximal enhancement achievable with QLs can be twice larger than for an homogeneous medium if the QLs occupy the intensity peaks of the resonance (Eq.~\ref{eq:F_equal_I}), leading to:

\begin{equation}
    F_{max}(\lambda_0) \approx \frac{16}{\ln(1/R_f)}.
    \label{eq:Fmax}
\end{equation}

\section{Numerical simulations}

Transfer matrix method (TMM)~\cite{katsidis_general_2002} and scattering matrix method~\cite{yuffa_object-oriented_2012} are two common ways to calculate the distribution of the electric field in planar thin-film stacks.
In this work, we use an in-house TMM algorithm that computes the absorption $A(\lambda)$ as well as the normalized field intensity $I(\lambda,z)$.

\subsection{Single resonant mode}

\begin{figure*}[t]
    \includegraphics[width=170mm]{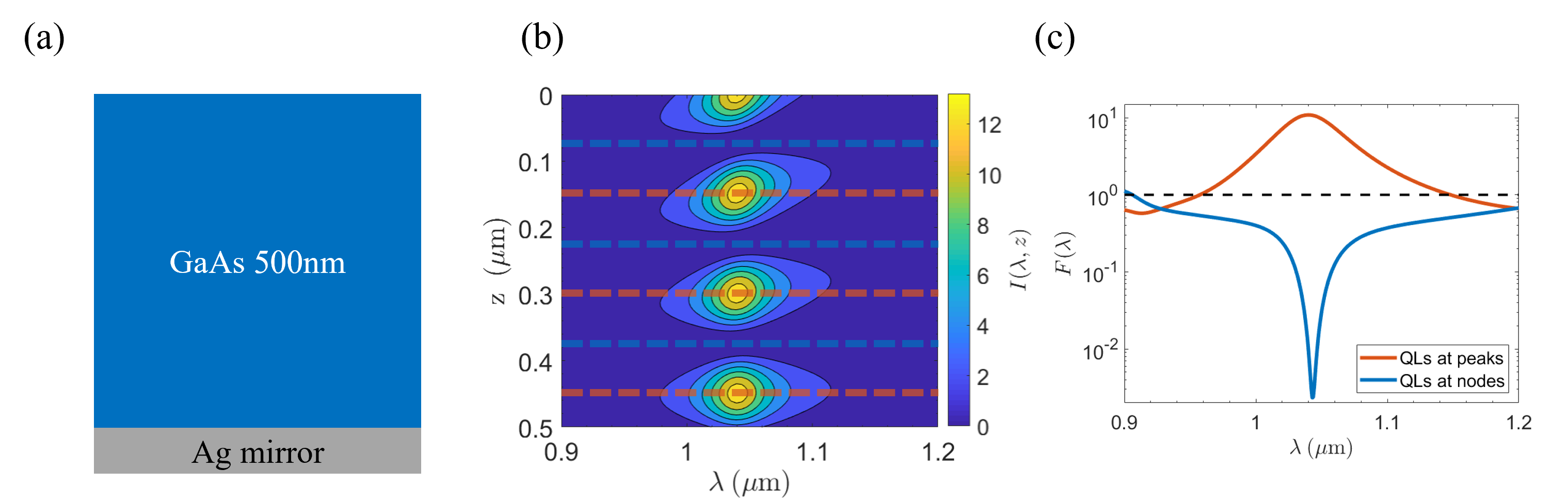}
    \caption{\label{fig:Optim_Unoptim} 
    (a) Structure simulated for a single resonant mode: a 500 nm-thick GaAs slab with a silver back mirror. (b) Normalized field intensity $I(\lambda,z)$ in the GaAs slab as a function of the wavelength and vertical position. We observe a single mode centered around $\lambda_0=1.040 \ \mathrm{\mu m}$. The contour lines correspond to multiples of 2. Two cases are considered. In the first, the QLs are inserted at the intensity peak of the field (red dashed lines). In the second, the QLs are inserted at the nodes of the field (blue dashes lines). (c) Average absorption enhancement $F$ in the QLs as a function of the wavelength, in log scale. The single pass absorption $F=1$ is represented with a dashed black line. When the QLs are at the peak positions (red line), the absorption enhancement at resonance is $F(\lambda_0) \approx 13$. When the QLs are at the nodes (blue line), the absorption at resonance is reduced by more than 2 orders of magnitude relative to single-pass absorption.
        }
\end{figure*}

We first illustrate the impact of the position of the QLs on a simple and general example with a single resonance.
We consider a 500 nm-thick GaAs\cite{palik_handbook_1998} slab on a silver\cite{jiang_realistic_2016} mirror (Fig.~\ref{fig:Optim_Unoptim}(a)), forming a FP cavity with air as the incident medium.
We calculate the normalized field intensity $I(\lambda,z)$ in the wavelength region $[0.9 \ \mathrm{\mu m}, 1.2 \ \mathrm{\mu m}]$, beyond the bandgap of GaAs (Fig.~\ref{fig:Optim_Unoptim}(b)).
We observe a single FP mode in that region, at a resonance wavelength $\lambda_0=1.040 \ \mathrm{\mu m}$.
This mode has 4 intensity peaks (including one at the surface) separated by a distance $\lambda_0/2n \approx 150 \ \mathrm{nm}$, with an intensity $I_{max} = 13.2$ at each peak.

To validate our model, we calculate the maximum light path enhancement achievable in this structure, using Eq.\ref{eq:Fmax}.
The front reflection at the GaAs/air interface is $R_f \approx 0.3$, leading to $F_{max}(\lambda_0) \approx 13.3$, in very good agreement with the value calculated for $I_{max}$.
This is the absorption enhancement we can expect for QLs positioned at the resonance peaks (assuming a small absorption from the QLs), twice higher than what can be achieved in homogeneous media.

We now insert three 1 nm-thick QLs in the slab, with a small absorption coefficient $\alpha_Q=10^4 \ \mathrm{cm^{-1}}$ independent of the wavelength.
We consider two cases (Fig.~\ref{fig:Optim_Unoptim}(b)). In the first case, the QLs are inserted at the intensity peaks of the resonant mode (red dashed lines). In the second case, the QLs are inserted at the nodes of the resonant mode (blue dashed lines).

We simulate both configurations and obtain the mean absorption enhancement in the QLs from Eq.~\ref{eq:mean_enhancement} (Fig.~\ref{fig:Optim_Unoptim}(c)).
When the QLs are inserted at the peaks of the mode, the mean absorption enhancement at resonance is $F(\lambda_0) = 12.93$, almost as large as the maximum field enhancement in absence of QLs (it is slightly smaller due to the absorption of the QLs).
However, when the QLs are inserted at the nodes of the mode, the absorption at resonance is almost completely suppressed, less than 1\% of the single-pass absorption.
Therefore, the position of the QLs has a tremendous impact on their absorption. In particular, an optical resonance does not systematically imply an absorption increase relative to single-pass absorption.
We note that the configuration where the QLs are inserted at the nodes may be useful for applications where it is beneficial to suppress absorption or emission at a specific wavelength.

\subsection{Multiple resonant modes}

\begin{figure*}[t]
    \centering
    \includegraphics[width=\linewidth]{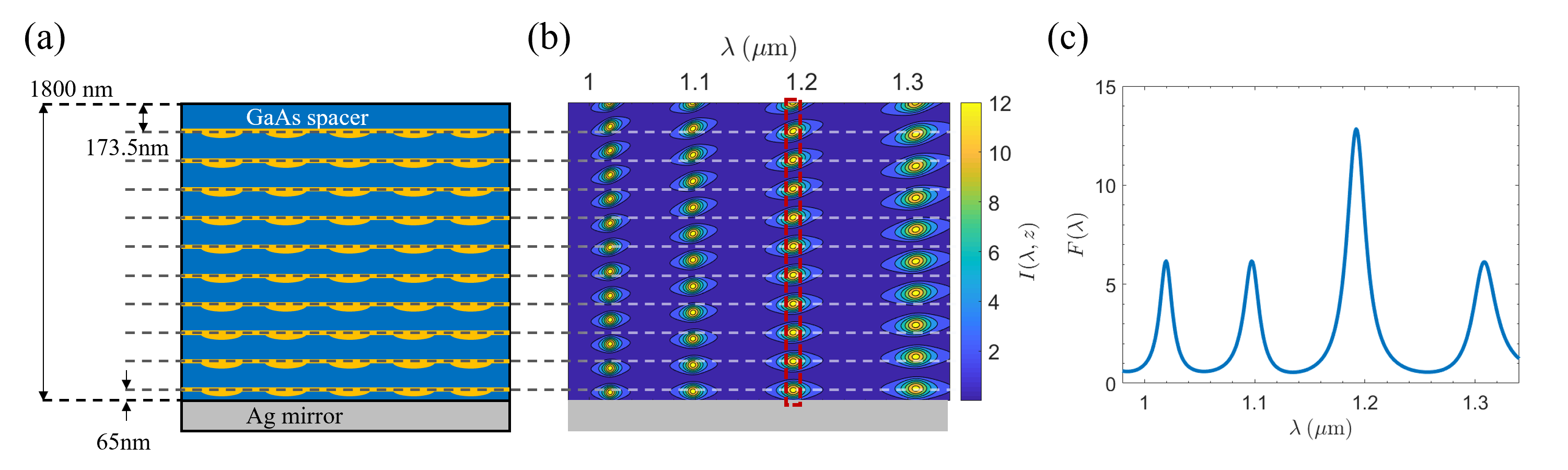}
    \caption{\label{fig:theoretical}
    (a) Structure simulated for several resonant modes: a 1800 nm-thick GaAs layer with 10 QLs on a silver back mirror. 
    (b) Normalized field intensity $I(\lambda,z)$ in the slab as a function of the wavelength and position. The contour lines correspond to multiples of 2. Four resonant modes are observed, with many lobes indicating high-order FP modes. The QLs are inserted at the peaks of the resonant mode centered around $\lambda_0 = 1192 \mathrm{nm}$. 
    (c) Light pass enhancement $F(\lambda)$ in the device. At the optimized resonance peak, the absorption enhancement is over 12, similar to the intensity of the field, while it is only slightly over 6 (about half) for the non-optimized resonances.
    }
\end{figure*}

We now consider a 1.8 \textmu m-thick GaAs slab on a silver mirror, for which we simulate the normalized field intensity (Fig.~\ref{fig:theoretical}(a,b))). 
In the spectral region [0.98 \textmu m, 1.34 \textmu m], we identify 4 resonant modes with many lobes. We focus on the resonant mode centered around $\lambda = 1.192 \ \mathrm{\mu m}$, which shows 10 peaks (not counting the one at the surface).
Since the peak intensity only depends on the front surface reflection (Eq.\ref{eq:Fmax}) and not, for example, on the absorber thickness, the maximum intensity is the same as in the previous case for all resonant modes, with $I_{max} \approx 13.3$.

By introducing 10 QLs into the slab, one at each peak, we can fully benefit from the resonant absorption of this resonant mode, leading to a light path enhancement over 12 (Fig.~\ref{fig:theoretical}(c)).
For the three non-optimized resonant modes, the light path enhancement is only slightly over 6, or about half the maximum value.
This is because we can consider the QLs are positioned randomly relative to the peaks of the non-optimized resonances. As a result, the light path enhancement $F$ takes the average value of the field enhancement $I_{max}/2$.
Therefore, by optimizing the position of QLs, we can increase the absorption by a factor 2 compared to a random distribution.

\section{Experimental demonstration}


In this section, we validate our theoretical framework and confirm experimentally the numerical observations of Fig.~\ref{fig:theoretical}(c).
To do so, we fabricate a structure similar to the one in Fig.~\ref{fig:theoretical}(a), with InAs QD layers as the QLs, and measure the absorption enhancement for optimized and unoptimized FP modes using photoluminescence.

\subsection{Sample fabrication}

\begin{figure}
    \centering
    \includegraphics[width=.8\linewidth]{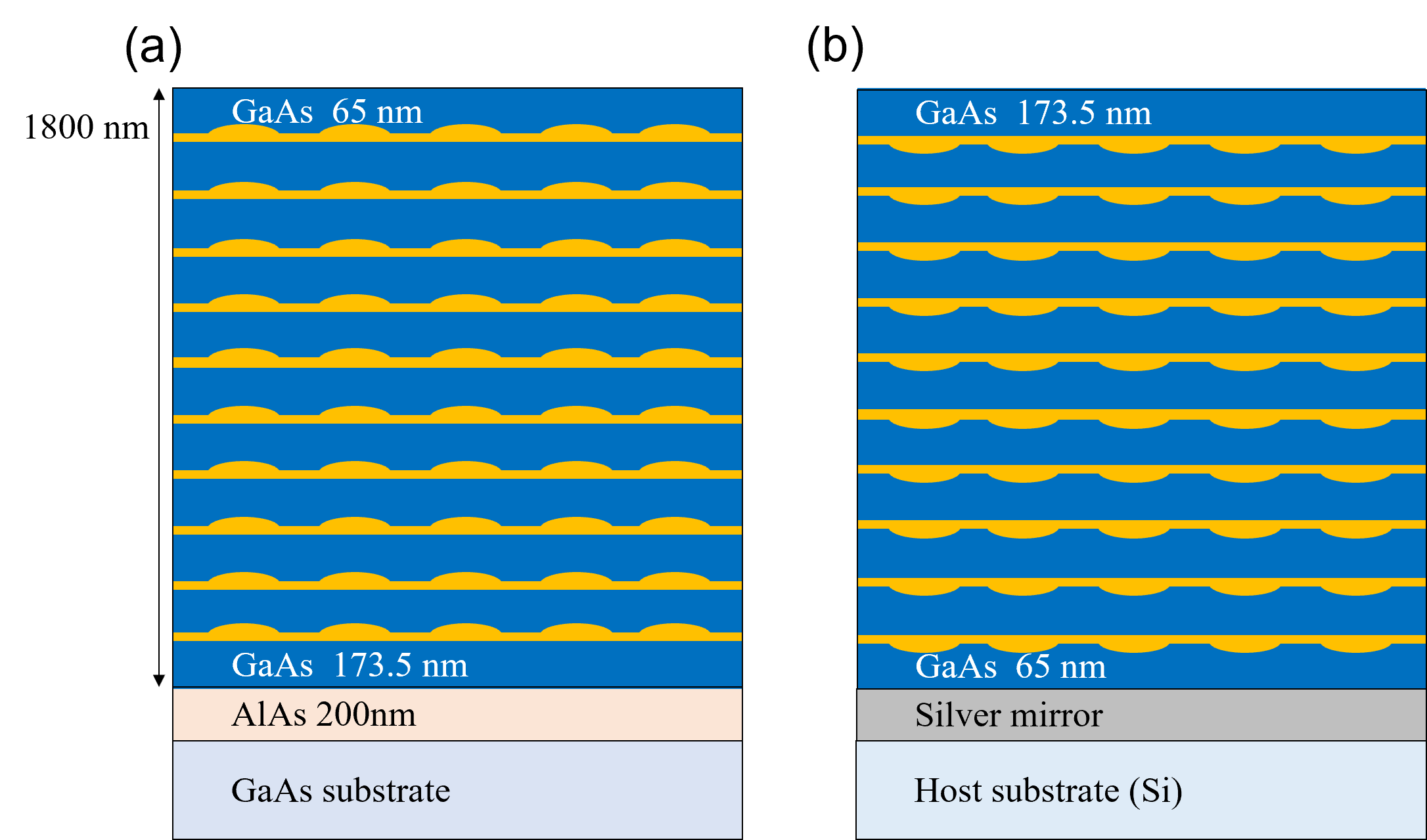}
    \caption{Samples fabricated to measure the influence of the position of QLs on  resonant absorption.
    (a) Reference (as-grown) sample: 1800 nm-thick GaAs host with 10 regularly spaced InAs QD layers, grown on a GaAs substrate, separated by an AlAs etch stop layer.
    (b) FP sample obtained after silver evaporation, bonding and substrate removal. The optical stack consists of the reversed 1800 nm thick GaAs/InAs slab on a mirror.
    }
    \label{fig:Samples}
\end{figure}

We grow a 1.8 \textmu m-thick GaAs layer with 10 layers of InAs QDs on a GaAs (001) double polished substrate by molecular beam epitaxy (MBE).
First, a GaAs buffer layer of about 300 nm is grown at a temperature of 570${\mathrm{^\circ C}}$, followed by a 200nm-thick AlAs release layer.
Then, the substrate temperature is lowered to 482${\mathrm{^\circ C}}$ to grow the 1800 nm-thick GaAs/InAs multilayer. A 1.8 monolayer (ML)-thick InAs QD layer is grown after every 173.5 nm of GaAs, and the last QD layer is capped with 65 nm of GaAs. 
The InAs QD layers are grown at a rate of 0.04 ML/s under As pressure (${1\times10^{-4}\,\mathrm{Pa}}$), while the GaAs spacer layers are grown at a rate of 1.0 ML/s under As pressure (${3\times10^{-4}\,\mathrm{Pa}}$).
From high-resolution X-ray diffraction, we estimate the spacer thickness to be $172 \pm 4 \ \mathrm{nm}$, in perfect agreement with the designed thickness.
This as-grown stack constitutes the reference (ref) sample (Fig.~\ref{fig:Samples}(a)).

This sample is then processed to obtain the Fabry-Perot (FP) sample (Fig.~\ref{fig:Samples}(b)).
A silver mirror (100 nm) is deposited on the front of the epitaxial stack, and covered with gold (300 nm) for protection, using thermal evaporation. The sample is then bonded to a silicon wafer with epoxy resin.
The GaAs substrate is etched first in a $\mathrm{NH_3OH:H_2O_2:H_2O}$ (1:1:2) solution, and continued in citric acid:$\mathrm{H_2O_2}$ (1:5), which is slower but more selective with AlAs.
Finally, the AlAs layer is etched in a buffered HF solution.

\subsection{Expressions for the absorption}

The absorption of the reference sample is obtained from the Beer-Lambert law (neglecting the impact of the AlAs layer and the reflection from the back surface):

\begin{equation}
        A_{ref}(E) = (1-R_{f}(E)) \left[1-\exp(-M \alpha_Q(E) d_Q)\right],
        \label{eq:A_ref_exp}
\end{equation}
    
\noindent where $R_f$ is the front surface reflection of the sample and $E = h c/\lambda$ is the photon energy, with $h$ the Planck constant. Since the absorption from the QDs is small (reflection-transmission measurements indicate that single-pass absorption is less than 1\%), it can be approximated as
    
\begin{equation}
     A_{ref}(E) \approx (1-R_{f}(E)) M \alpha_Q(E) d_Q,
     \label{eq:A_ref_lin}
\end{equation}

\noindent while the absorption of the FP sample writes, according to Eq.~\ref{eq:A_lambda_Fm}:

\begin{equation}
    A_{FP}(E) = \alpha_Q(E) d_Q \sum_{m=1}^M F_m(E).
    \label{eq:A_FP}
\end{equation}

\subsection{Absorption enhancement from photoluminescence}

Since absorption is very small, especially in the reference sample, reflection-transmission measurements do not enable an accurate determination of the absorption enhancement.
Moreover, these measurements include the absorption from other layers (free-carrier absorption in the GaAs substrate and parasitic absorption in the mirror), which can be difficult to separate from that of the QDs.
To measure the absorption enhancement in weakly absorbing media, typical methods include EQE~\cite{tamaki_spectrally_2014} and waveguide absorption~\cite{yoshikawa_observation_2018}, which respectively require electrical contacts and specific fabrication processes.
Photoluminescence (PL), which consists in measuring the spectral photon flux emitted by a sample under laser illumination, is another approach to determine the absorption coefficient of weakly absorbing media. 
It relies on Kirchhoff's law to determine absorption from emission, and offers the advantage of requiring no additional processing.
It has been applied successfully to measure the absorption coefficient near the bandgap in materials such as silicon~\cite{daub_ultralow_1995} or thin-film CIGS~\cite{rey_absorption_2018}.
Although the determination of the absorption coefficient requires the absolute value of the PL flux~\cite{rey_absorption_2018}, we show in the following that the absorption enhancement $F(E)$ can be determined from the ratio of PL spectral intensities alone.

The photon flux spectrum emitted by a semiconductor per unit solid angle can be described by the generalized Planck law \cite{wurfel_chemical_1982}:

\begin{equation}
    \Phi(E) = A(E) \frac{2 E^2}{h^3c^2}\left[ \exp\left(\frac{E-\Delta\mu}{k_B T}\right) -1 \right]^{-1},
\end{equation}

\noindent where $\Delta\mu$ is the quasi-Fermi level splitting (QFLS) between electrons and holes and $T$ is the lattice temperature.
This expression can be extended to describe the emission of multiple QLs provided these layers are identical and all share the same $\Delta \mu$.
When the incident illumination is not too strong, we can consider the Boltzmann approximation:

\begin{equation}
    \Phi(E) \approx A(E) \frac{2 E^2}{h^3c^2} \exp\left(-\frac{E}{k_B T}\right) \exp\left(\frac{\Delta\mu}{k_B T}\right).
    \label{eq:generalized_Planck_law_MB}
\end{equation}

The ratio $r$ between the PL spectra from the sample after processing $\Phi_{FP}$ and before processing $\Phi_{ref}$ thus takes the expression:

\begin{equation}
    r(E) = \frac{\Phi_{FP}(E)}{\Phi_{ref}(E)}=\frac{A_{FP}(E)}{2 A_{ref}(E)} \exp\left( \frac{\Delta\mu_{FP}-\Delta\mu_{ref}}{k_B T} \right),
    \label{eq:PL_ratio_A_ratio}
\end{equation}

\noindent where the factor 2 accounts for the emission from the back of the substrate in the reference case.
We assume the illumination intensity is sufficiently low so that we can neglect band filling and other non-linear effects, and absorption is obtained from Eqs.~\ref{eq:A_ref_lin}-\ref{eq:A_FP}.
This is justified because $r$ is relatively independent of the incident power (see Fig.~\ref{fig:intensity_dependent_PL}). 
The PL ratio becomes:

\begin{equation}
    r(E) = \frac{F(E)}{2(1-R_{f}(E))} \exp\left( \frac{\mu_{FP}-\mu_{ref}}{k_B T} \right).
    \label{eq:PL_ratio_F}
\end{equation}

The energy dependence mostly comes from the variations of $F$ (the front reflection is almost constant in the spectral region considered). Therefore, the PL ratio $r$ is proportional to the light path enhancement $F$ and is independent of the exact value of the absorption coefficient.

\subsection{Experimental validation}

\begin{figure*}[t]
    \centering
    \includegraphics[width=170mm]{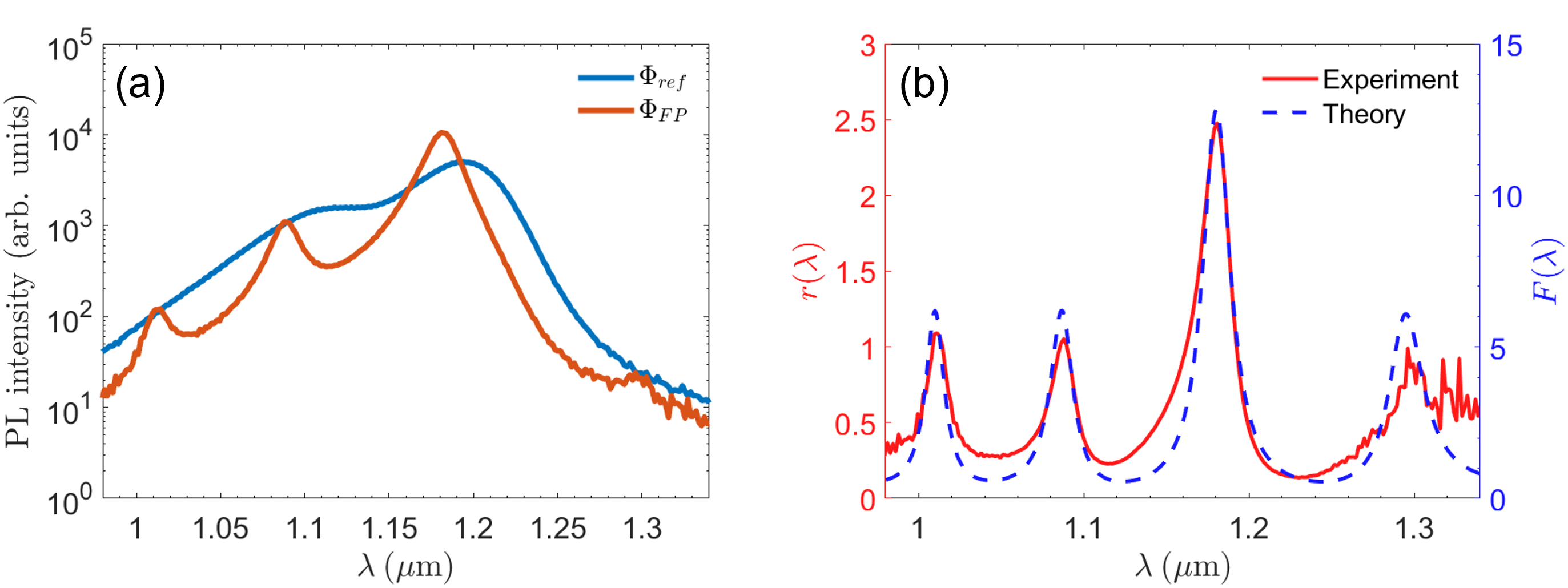}
    \caption{
    (a) Room-temperature PL spectra measured on reference ($\Phi_{ref}$, blue line) and FP ($\Phi_{FP}$, red line) samples, for a incident laser wavelength $\lambda = 782 \ \mathrm{nm}$ and power $P = 100 \ \mathrm{W.cm^{-2}}$ (log scale).
    (b) Ratio $r$ of the PL spectra shown in (a) (red line, left axis) and light path enhancement $F$ simulated for a 1780 nm-thick GaAs slab with 10 QD layers positioned to maximize the absorption for the FP resonance at 1.18 \textmu m  (dashed blue line, right axis, different scale). We observe a good agreement between theory and experiment.
    }
    \label{fig:PL_ratio}
\end{figure*}

Fig.\ref{fig:PL_ratio}(a) shows room-temperature PL spectra obtained for the reference and FP samples (Fig.\ref{fig:Samples}), for the same laser excitation ($\lambda = 782 \ \mathrm{nm}$, $P = 100 \ \mathrm{W.cm^{-2}}$ with a spot diameter of 100 \textmu m).
The excitation energy is above the bandgap of GaAs, so the carriers are mainly generated in the host.
Some of these carriers recombine radiatively in the QDs, leading to the measured PL signal.

The PL of the reference sample $\Phi_{ref}$ shows two peaks, the main one around $\lambda = 1210 \ \mathrm{nm}$ corresponding to the fundamental transition in the QDs.
The second peak can be attributed to an excited state or to the size distribution of the QDs.
We calculate the ratio $r$ between both PL spectra (Fig.~\ref{fig:PL_ratio}(b), red line).
We observe 4 peaks at 1011 nm, 1088 nm, 1181 nm and 1294 nm, with a significantly higher intensity for the third peak.

\begin{figure}[t]
    \centering
    \includegraphics[width=.6\linewidth]{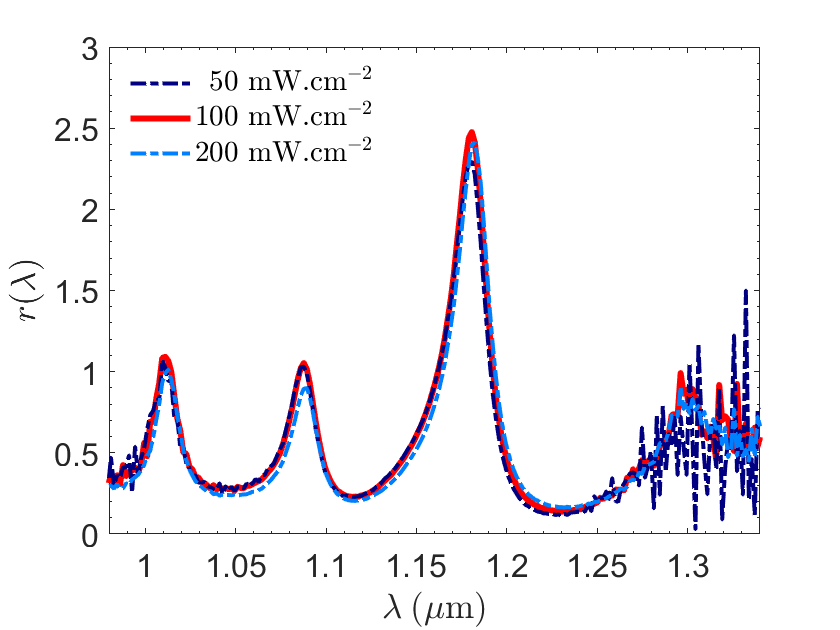}
    \caption{PL ratio spectra $r(\lambda)$ for different laser intensities.
    For laser powers twice smaller and larger than $100 \ \mathrm{W.cm^{-2}}$ (the power considered in Fig.~\ref{fig:PL_ratio}), the changes in the PL ratio are small. In particular the resonant peak at 1192 nm always shows a much larger ratio than the other peaks, confirming the impact of the position of the QLs on absorption.}
    \label{fig:intensity_dependent_PL}
\end{figure}

To compute the light path enhancement $F$ (Fig.~\ref{fig:PL_ratio}(b), blue dashed line), we adjust the total stack thickness in the model to 1780 nm.
The small difference (1\% compared to the designed structure) can be attributed to the uncertainty in the growth rate.
We observe a very good match in the peak position between both curves in Fig.~\ref{fig:PL_ratio}(b), confirming that the oscillations in the PL ratio originate from the FP resonances.
Furthermore, the relative intensity of the peaks of $r$ agrees with those of $F$: the peak for which the QD position is optimized shows an enhancement more than twice that of the other peaks.
Also, $r$ is almost independent of the laser intensity (see Fig.~\ref{fig:intensity_dependent_PL}), confirming that this difference in the relative intensity of the FP peaks cannot be attributed to non-linear effects in the QDs.
Overall, we verify experimentally that the position of multiple QLs can have a significant impact on their absorption, allowing higher absorption enhancements than with homogeneous media.

We notice two differences in the shape of $r$ compared to $F$: an asymmetry of the main PL ratio peak, which is broader than expected on the low-wavelength side, and a ratio larger than 2 between the peaks. 
These could both originate from spatial variations of the QD distribution (the PL spectra of the reference and FP samples were measured in positions separated by about 1 mm on the original wafer).

\subsection{Difference in quasi-Fermi level splitting between reference and FP samples}
\label{s_factor5}

We note that the scale for the PL ratio $r$ and the light path enhancement $F$ in Fig.~\ref{fig:PL_ratio}(b) are different, with $F/r \approx 5$.
Using Eq.~\ref{eq:PL_ratio_F} with $R_f \approx 0.30$, we can estimate the difference in QFLS between both samples, for the same incident power:

\begin{equation}
    \Delta\mu_{ref}-\Delta\mu_{FP} \approx 33 \ \mathrm{meV}.
    \label{eq:mu_difference_experimental}
\end{equation}

The absorption at the laser wavelength is almost the same for both samples (63\% and 62\%, respectively), so the generated carrier density can be considered identical.
Three distinct mechanisms may contribute to explaining why the QFLS in the FP sample is smaller than in the reference sample (Eq.~\ref{eq:mu_difference_experimental}).

A first explanation for the lower QFLS in the FP sample is the difference in the back interface (Fig.~\ref{fig:Samples}). For the reference sample, the AlAs layer acts as a barrier for the carriers, preventing recombinations. However, for the FP sample, the silver interface leads to parasitic recombinations, reducing the carrier density in the sample and therefore the QFLS.
This could be prevented by adding an Al(Ga)As barrier between the epitaxial layer and the mirror in the FP sample.

A second possible contribution to QFLS reduction is sample quality. 
Indeed, the FP sample was slightly degraded after processing (apparition of pinholes), which can reduce the QFLS due to increased parasitic recombinations.

Even if we could neglect the two first contributions, the QFLS in the FP sample should be smaller than in the reference sample. Indeed, if we assume identical generation and non-radiative recombinations in both samples, then, from detailed balance considerations, radiative recombinations must also be identical: $\int \Phi_{FP}(E) dE = \int \Phi_{ref}(E) dE$.
However, the absorption differs significantly between both samples. 
As the absorption is generally larger in the FP sample, especially close to the emission peak, the QFLS must be smaller to ensure equal radiative recombinations.

\section{Conclusion}

We have highlighted the crucial impact of the position of multiple QLs within a host material on their resonant absorption.
We introduced a formalism to calculate the light absorption enhancement factor $F(\lambda)$ for a given distribution of the QLs, showing that this enhancement can be up to twice larger than in homogeneous media when the QLs are placed at the peaks of the field intensity.
We then observed numerically how adjusting the position of the QLs can lead to strongly increasing or suppressing the absorption of a resonant mode. In particular we showed that an optimal position enables an absorption enhancement twice larger than a random position.
Finally, we fabricated devices with multiple InAs QD layers. We presented a method to determine absorption enhancement from PL spectra measured on samples with and without a resonant cavity, from which we confirmed quantitatively the larger absorption enhancement for an optimized QL position.

This work is of interest for all application where increasing absorption in QLs can lead to an improvement in device performance.
It is particularly relevant to increase current generation in high-efficiency PV concepts such as intermediate-band solar cells.
It is important to note that the benefits of resonant absorption highlighted here can only be obtained in thin devices.
The formalism we developed can be extended to optimize broadband multi-resonant absorption~\cite{yu_fundamental_2010}.
Although we focused here on the absorption of interband transitions within the QLs, this approach could also be used to increase the absorption of intersubband transitions for intermediate-band solar cells~\cite{tamaki_spectrally_2014}.
Finally, it can be adapted to thin superlattice stacks, where several QLs are grown very close to each other~\cite{harada_hot-carrier_2019}.

\appendix    

\section{Maximum enhancement with quantum layers}
\label{sec:AppendixA}

The temporal coupled-mode theory framework is helpful to analyze the absorption spectrum of a resonant mode in a homogeneous cavity~\cite{haus_waves_1984,yu_fundamental_2010,collin_nanostructure_2014}.
In this section, we consider a frequency description with

\begin{equation}
    \omega = \frac{2 \pi c}{n \lambda},
\end{equation}

\noindent where $c$ is the speed of light in vacuum and $n$ is the refractive index inside the cavity.
Assuming the full-width at half maximum (FWHM) of a resonance $\Delta\omega$ is small compared to its resonant frequency $\omega_0$, the absorption spectrum follows a Lorentzian function~\cite{haus_waves_1984,yu_fundamental_2010,collin_nanostructure_2014}: 

\begin{equation}
    A(\omega) = \frac{4 \gamma_r \gamma_{nr}}{(\omega - \omega_0)^2 + (\gamma_r + \gamma_{nr})^2},
\end{equation}

\noindent where $\gamma_r$ is the radiative decay corresponding to the coupling of the cavity with the outside environment, and $\gamma_{nr}$ is the non-radiative decay due to absorption inside the cavity.

Assuming the mode is confined within the cavity, the non-radiative decay can be expressed as a function of the absorption coefficient $\alpha$\cite{yu_fundamental_2010}:

\begin{equation}
   \gamma_{nr} = \frac{\alpha c}{2 n}.
   \label{eq:gamma_nr}
\end{equation}

In a FP cavity, assuming a perfect back reflector, the radiative decay can be approximated by~\cite{saleh_fundamentals_2007,vandamme_ultrathin_2015}:

\begin{equation}
   \gamma_r = \frac{c}{4 n d} \ln(1/R_f),
   \label{eq:gamma_r}
\end{equation}

\noindent where $R_f$ is the front surface reflection.
If absorption is small, we can assume $\gamma_{nr} \ll \gamma_r$, such that:

\begin{equation}
    A(\omega) \approx \frac{4 \gamma_r \gamma_{nr}}{(\omega - \omega_0)^2 + \gamma_r^2},
\end{equation}

\noindent which we can also write:

\begin{equation}
    A(\omega) \approx \frac{4 \gamma_{nr}}{\gamma_{r}} \frac{1}{1 + \left( \frac{\omega - \omega_0}{\gamma_r} \right)^2}.
\end{equation}

The absorption at resonance $A(\omega_0)$, FWHM $\Delta\omega$ and absorption cross-section $\sigma = \int{A(\omega) d\omega}$ then take the expressions:

\begin{align}
    A(\omega_0) &\approx \frac{4 \gamma_{nr}}{\gamma_r} \label{eq:A_omega_0},\\
    \Delta\omega &\approx 2 \gamma_r \label{eq:FWHM},\\
    \sigma &\approx 4 \pi \gamma_{nr} \label{eq:cross_sec}.
\end{align}

Combining Eqs.\ref{eq:A_omega_0}, \ref{eq:gamma_nr} and \ref{eq:gamma_r}, the absorption at resonance can be written:

\begin{equation}
    A(\omega_0) \approx \alpha d \frac{8}{\ln(1/R_f)},
\end{equation}

\noindent and we can define the absorption enhancement at resonance relative to single-pass absorption:

\begin{equation}
    F(\omega_0) \approx \frac{8}{\ln(1/R_f)}.
\end{equation}

The higher the front reflection $R_f$, the stronger the absorption enhancement at resonance. However, the absorption bandwidth gets narrower as the radiative decay $\gamma_r$ decreases (Eq.~\ref{eq:FWHM}). As a result, the absorption cross-section is independent of $\gamma_r$ (Eq.\ref{eq:cross_sec}), meaning that increasing $R_f$ does not improve broadband absorption.
We note that in the general case (not assuming $\gamma_{nr} \ll \gamma_r$), the absorption can be written as a function of $F$, following:

\begin{equation}
    A(\omega_0) = \frac{F(\omega_0) \alpha d}{\left(1 +\frac{F(\omega_0) \alpha d}{4}\right)^2}.
\end{equation}

\subsection*{Disclosures}
The authors declare having no conflicts of interest.

\subsection*{Acknowledgments}

This work is supported by the New Energy and Industrial Technology Development Organization (NEDO) and the Ministry of Economy, Trade and Industry (METI), Japan (Grant No. 20000934-0). This work is also supported by the Japanese Society for the Promotion of Science (JSPS) KAKENHI (Grant No. 19H02541).

The authors would like to thank Stéphane Collin (C2N) and Jean-François Guillemoles (CNRS IPVF) for the valuable discussions.

\section*{Data Availability}

The data that support the findings of this study are available from the corresponding author upon reasonable request.

\section*{References}

\bibliography{FP_QD}
\bibliographystyle{spiejour}   


\vspace{2ex}\noindent\textbf{Maxime Giteau} is a postdoctoral researcher at RCAST, The University of Tokyo, Japan. He received his MSc degrees in engineering and solid state physics, respectively from Supaero and Paul Sabatier University, France, in 2016, and his PhD from the University of Tokyo in 2020. His current research interests include theoretical and practical aspects of light trapping as well as high-efficiency photovoltaic concepts such as hot-carrier solar cells.

\vspace{2ex}\noindent\textbf{Yusuke Oteki} is a doctoral student at RCAST, The University of Tokyo, Japan. He received his master’s degree in engineering from the University of Tokyo in 2020. His current research interests include light trapping in thin-film solar cells, in particular its implementation in high-efficiency photovoltaic concepts such as quantum dot intermediate band solar cells.

\vspace{2ex}\noindent\textbf{Yoshitaka Okada} received his bachelor’s degree in electronic and electrical engineering from the University of London in 1984 and his master’s degree and PhD from the University of Tokyo in 1987 and 1990, respectively. He is currently a professor at the RCAST of the University of Tokyo. His research interests include epitaxial growth of III-V-N dilute nitrides, 2D dichalcogenides, and low-dimensional quantum nanostructures for applications to high-efficiency and thin-film solar cells.

\vspace{1ex}
\noindent Biographies of the other authors are not available.

\listoffigures

\end{spacing}
\end{document}